
\message
{JNL.TEX version 0.92 as of 4/24/89.  Using CM fonts.}

\catcode`@=11
\expandafter\ifx\csname inp@t\endcsname\relax\let\inp@t=\input
\def\input#1 {\expandafter\ifx\csname #1IsLoaded\endcsname\relax
\inp@t#1%
\expandafter\def\csname #1IsLoaded\endcsname{(#1 was previously loaded)}
\else\message{\csname #1IsLoaded\endcsname}\fi}\fi
\catcode`@=12

\font\twelverm=cmr12			\font\twelvei=cmmi12
\font\twelvesy=cmsy10 scaled 1200	\font\twelveex=cmex10 scaled 1200
\font\twelvebf=cmbx12			\font\twelvesl=cmsl12
\font\twelvett=cmtt12			\font\twelveit=cmti12
\font\twelvesc=cmcsc10 scaled 1200	\font\twelvesf=cmss12
\skewchar\twelvei='177			\skewchar\twelvesy='60


\def\twelvepoint{\normalbaselineskip=12.4pt plus 0.1pt minus 0.1pt
  \abovedisplayskip 12.4pt plus 3pt minus 9pt
  \belowdisplayskip 12.4pt plus 3pt minus 9pt
  \abovedisplayshortskip 0pt plus 3pt
  \belowdisplayshortskip 7.2pt plus 3pt minus 4pt
  \smallskipamount=3.6pt plus1.2pt minus1.2pt
  \medskipamount=7.2pt plus2.4pt minus2.4pt
  \bigskipamount=14.4pt plus4.8pt minus4.8pt
  \def\rm{\fam0\twelverm}          \def\it{\fam\itfam\twelveit}%
  \def\sl{\fam\slfam\twelvesl}     \def\bf{\fam\bffam\twelvebf}%
  \def\mit{\fam 1}                 \def\cal{\fam 2}%
  \def\sc{\twelvesc}		   \def\tt{\twelvett}
  \def\sf{\twelvesf}
  \textfont0=\twelverm   \scriptfont0=\tenrm   \scriptscriptfont0=\sevenrm
  \textfont1=\twelvei    \scriptfont1=\teni    \scriptscriptfont1=\seveni
  \textfont2=\twelvesy   \scriptfont2=\tensy   \scriptscriptfont2=\sevensy
  \textfont3=\twelveex   \scriptfont3=\twelveex  \scriptscriptfont3=\twelveex
  \textfont\itfam=\twelveit
  \textfont\slfam=\twelvesl
  \textfont\bffam=\twelvebf \scriptfont\bffam=\tenbf
  \scriptscriptfont\bffam=\sevenbf
  \normalbaselines\rm}


\def\beginlinemode{\endmode
  \begingroup\parskip=0pt \obeylines\def\\{\par}\def\endmode{\par\endgroup}}
\def\beginparmode{\endmode
  \begingroup \def\endmode{\par\endgroup}}
\let\endmode=\par
{\obeylines\gdef\
{}}
\def\singlespace{\baselineskip=\normalbaselineskip}

\def\oneandahalfspace{\baselineskip=\normalbaselineskip
  \multiply\baselineskip by 3 \divide\baselineskip by 2}
\def\doublespace{\baselineskip=\normalbaselineskip \multiply\baselineskip by 2}

\newcount\firstpageno
\firstpageno=2
\footline={\ifnum\pageno<\firstpageno{\hfil}\else{\hfil\twelverm\folio\hfil}\fi}
\def\toppageno{\global\footline={\hfil}\global\headline
  ={\ifnum\pageno<\firstpageno{\hfil}\else{\hfil\twelverm\folio\hfil}\fi}}
\let\rawfootnote=\footnote		
\def\footnote#1#2{{\rm\singlespace\parindent=0pt\parskip=0pt
  \rawfootnote{#1}{#2\hfill\vrule height 0pt depth 6pt width 0pt}}}
\def\raggedcenter{\leftskip=4em plus 12em \rightskip=\leftskip
  \parindent=0pt \parfillskip=0pt \spaceskip=.3333em \xspaceskip=.5em
  \pretolerance=9999 \tolerance=9999
  \hyphenpenalty=9999 \exhyphenpenalty=9999 }


\hsize=6truein
\hoffset=.3truein
\vsize=8truein
\voffset=.3truein
\parskip=\medskipamount
\def\\{\cr}
\twelvepoint		
\doublespace		
\overfullrule=0pt	

\def\title			
  {\null\vskip 3pt plus 0.2fill
   \beginlinemode \doublespace \raggedcenter \bf}

\def\author			
  {\vskip 3pt plus 0.2fill \beginlinemode
   \singlespace \raggedcenter\sc}

\def\affil			
  {\vskip 3pt plus 0.1fill \beginlinemode
   \oneandahalfspace \raggedcenter \sl}

\def\abstract			
  {\vskip 3pt plus 0.3fill \beginparmode
   \oneandahalfspace ABSTRACT: }

\def\endpage{\vfill\eject}

\def\endtitlepage		
  {\endpage			
   \body}

\def\body			
  {\beginparmode}		

\def\head#1{			
  \goodbreak\vskip 0.5truein	
  {\immediate\write16{#1}
   \raggedcenter \uppercase{#1}\par}
   \nobreak\vskip 0.25truein\nobreak}

\def\references			
  {\head{References}		
   \beginparmode
   \frenchspacing \parindent=0pt \leftskip=1truecm
   \parskip=8pt plus 3pt \everypar{\hangindent=\parindent}}

\gdef\refis#1{\item{#1.\ }}			

\gdef\journal#1, #2, #3, 1#4#5#6{		
    {\sl #1~}{\bf #2}, #3 (1#4#5#6)}		

\def\pr{\journal Phys. Rev., }

\def\pl{\journal Phys. Lett., }

\def\endreferences{\body}

\def\figurecaptions		
  {\endpage
   \beginparmode
   \head{Figure Captions}
}

\def\endpaper			
  {\endmode\vfill\supereject}

\def\tag#1$${\eqno(#1)$$}

\def\align#1$${\eqalign{#1}$$}

\def\aligntag#1$${\gdef\tag##1\\{&(##1)\cr}\eqalignno{#1\\}$$
  \gdef\tag##1$${\eqno(##1)$$}}

\def\endaligntag{}

\def\overset #1\to#2{{\mathop{#2}\limits^{#1}}}
\def\underset#1\to#2{{\let\next=#1\mathpalette\undersetpalette#2}}
\def\undersetpalette#1#2{\vtop{\baselineskip0pt
\ialign{$\mathsurround=0pt #1\hfil##\hfil$\crcr#2\crcr\next\crcr}}}


\catcode`@=11
\newcount\tagnumber\tagnumber=0

\immediate\newwrite\eqnfile
\newif\if@qnfile\@qnfilefalse
\def\write@qn#1{}
\def\writenew@qn#1{}
\def\w@rnwrite#1{\write@qn{#1}\message{#1}}
\def\@rrwrite#1{\write@qn{#1}\errmessage{#1}}

\def\taghead#1{\gdef\t@ghead{#1}\global\tagnumber=0}
\def\t@ghead{}

\expandafter\def\csname @qnnum-3\endcsname
  {{\t@ghead\advance\tagnumber by -3\relax\number\tagnumber}}
\expandafter\def\csname @qnnum-2\endcsname
  {{\t@ghead\advance\tagnumber by -2\relax\number\tagnumber}}
\expandafter\def\csname @qnnum-1\endcsname
  {{\t@ghead\advance\tagnumber by -1\relax\number\tagnumber}}
\expandafter\def\csname @qnnum0\endcsname
  {\t@ghead\number\tagnumber}
\expandafter\def\csname @qnnum+1\endcsname
  {{\t@ghead\advance\tagnumber by 1\relax\number\tagnumber}}
\expandafter\def\csname @qnnum+2\endcsname
  {{\t@ghead\advance\tagnumber by 2\relax\number\tagnumber}}
\expandafter\def\csname @qnnum+3\endcsname
  {{\t@ghead\advance\tagnumber by 3\relax\number\tagnumber}}

\def\equationfile{%
  \@qnfiletrue\immediate\openout\eqnfile=\jobname.eqn%
  \def\write@qn##1{\if@qnfile\immediate\write\eqnfile{##1}\fi}
  \def\writenew@qn##1{\if@qnfile\immediate\write\eqnfile
    {\noexpand\tag{##1} = (\t@ghead\number\tagnumber)}\fi}
}

\def\callall#1{\xdef#1##1{#1{\noexpand\call{##1}}}}
\def\call#1{\each@rg\callr@nge{#1}}

\def\each@rg#1#2{{\let\thecsname=#1\expandafter\first@rg#2,\end,}}
\def\first@rg#1,{\thecsname{#1}\apply@rg}
\def\apply@rg#1,{\ifx\end#1\let\next=\relax%
\else,\thecsname{#1}\let\next=\apply@rg\fi\next}

\def\callr@nge#1{\calldor@nge#1-\end-}
\def\callr@ngeat#1\end-{#1}
\def\calldor@nge#1-#2-{\ifx\end#2\@qneatspace#1 %
  \else\calll@@p{#1}{#2}\callr@ngeat\fi}
\def\calll@@p#1#2{\ifnum#1>#2{\@rrwrite{Equation range #1-#2\space is bad.}
\errhelp{If you call a series of equations by the notation M-N, then M and
N must be integers, and N must be greater than or equal to M.}}\else%
 {\count0=#1\count1=#2\advance\count1
by1\relax\expandafter\@qncall\the\count0,%
  \loop\advance\count0 by1\relax%
    \ifnum\count0<\count1,\expandafter\@qncall\the\count0,%
  \repeat}\fi}

\def\@qneatspace#1#2 {\@qncall#1#2,}
\def\@qncall#1,{\ifunc@lled{#1}{\def\next{#1}\ifx\next\empty\else
  \w@rnwrite{Equation number \noexpand\(>>#1<<) has not been defined yet.}
  >>#1<<\fi}\else\csname @qnnum#1\endcsname\fi}

\let\eqnono=\eqno
\def\eqno(#1){\tag#1}
\def\tag#1$${\eqnono(\displayt@g#1 )$$}

\def\aligntag#1\endaligntag
  $${\gdef\tag##1\\{&(##1 )\cr}\eqalignno{#1\\}$$
  \gdef\tag##1$${\eqnono(\displayt@g##1 )$$}}

\def\eqalignno#1{\displ@y \tabskip\centering
  \halign to\displaywidth{\hfil$\displaystyle{##}$\tabskip\z@skip
    &$\displaystyle{{}##}$\hfil\tabskip\centering
    &\llap{$\displayt@gpar##$}\tabskip\z@skip\crcr
    #1\crcr}}

\def\displayt@gpar(#1){(\displayt@g#1 )}

\def\displayt@g#1 {\rm\ifunc@lled{#1}\global\advance\tagnumber by1
        {\def\next{#1}\ifx\next\empty\else\expandafter
        \xdef\csname @qnnum#1\endcsname{\t@ghead\number\tagnumber}\fi}%
  \writenew@qn{#1}\t@ghead\number\tagnumber\else
        {\edef\next{\t@ghead\number\tagnumber}%
        \expandafter\ifx\csname @qnnum#1\endcsname\next\else
        \w@rnwrite{Equation \noexpand\tag{#1} is a duplicate number.}\fi}%
  \csname @qnnum#1\endcsname\fi}

\def\ifunc@lled#1{\expandafter\ifx\csname @qnnum#1\endcsname\relax}

\let\@qnend=\end\gdef\end{\if@qnfile
\immediate\write16{Equation numbers written on []\jobname.EQN.}\fi\@qnend}

\catcode`@=12

\catcode`@=11
\newcount\r@fcount \r@fcount=0
\newcount\r@fcurr
\immediate\newwrite\reffile
\newif\ifr@ffile\r@ffilefalse
\def\w@rnwrite#1{\ifr@ffile\immediate\write\reffile{#1}\fi\message{#1}}

\def\writer@f#1>>{}
\def\referencefile{
  \r@ffiletrue\immediate\openout\reffile=\jobname.ref%
  \def\writer@f##1>>{\ifr@ffile\immediate\write\reffile%
    {\noexpand\refis{##1} = \csname r@fnum##1\endcsname = %
     \expandafter\expandafter\expandafter\strip@t\expandafter%
     \meaning\csname r@ftext\csname r@fnum##1\endcsname\endcsname}\fi}%
  \def\strip@t##1>>{}}

\def\citeall#1{\xdef#1##1{#1{\noexpand\cite{##1}}}}
\def\cite#1{\each@rg\citer@nge{#1}}	
\def\reff#1{Reference [\cite{#1}]}
\def\each@rg#1#2{{\let\thecsname=#1\expandafter\first@rg#2,\end,}}
\def\first@rg#1,{\thecsname{#1}\apply@rg}	
\def\apply@rg#1,{\ifx\end#1\let\next=\relax
\else,\thecsname{#1}\let\next=\apply@rg\fi\next}
\def\citer@nge#1{\citedor@nge#1-\end-}	
\def\citer@ngeat#1\end-{#1}
\def\citedor@nge#1-#2-{\ifx\end#2\r@featspace#1 
  \else\citel@@p{#1}{#2}\citer@ngeat\fi}	
\def\citel@@p#1#2{\ifnum#1>#2{\errmessage{Reference range #1-#2\space is bad.}%
    \errhelp{If you cite a series of references by the notation M-N, then M and
    N must be integers, and N must be greater than or equal to M.}}\else%
 {\count0=#1\count1=#2\advance\count1
by1\relax\expandafter\r@fcite\the\count0,%
  \loop\advance\count0 by1\relax
    \ifnum\count0<\count1,\expandafter\r@fcite\the\count0,%
  \repeat}\fi}

\def\r@featspace#1#2 {\r@fcite#1#2,}	
\def\r@fcite#1,{\ifuncit@d{#1}
    \newr@f{#1}%
    \expandafter\gdef\csname r@ftext\number\r@fcount\endcsname%
                     {\message{Reference #1 to be supplied.}%
                      \writer@f#1>>#1 to be supplied.\par}%
 \fi%
 \csname r@fnum#1\endcsname}
\def\ifuncit@d#1{\expandafter\ifx\csname r@fnum#1\endcsname\relax}%
\def\newr@f#1{\global\advance\r@fcount by1%
    \expandafter\xdef\csname r@fnum#1\endcsname{\number\r@fcount}}

\let\r@fis=\refis			
\def\refis#1#2#3\par{\ifuncit@d{#1}
   \newr@f{#1}%
   \w@rnwrite{Reference #1=\number\r@fcount\space is not cited up to now.}\fi%
  \expandafter\gdef\csname r@ftext\csname r@fnum#1\endcsname\endcsname%
  {\writer@f#1>>#2#3\par}}

\def\ignoreuncited{
   \def\refis##1##2##3\par{\ifuncit@d{##1}%
     \else\expandafter\gdef\csname r@ftext\csname
r@fnum##1\endcsname\endcsname%
     {\writer@f##1>>##2##3\par}\fi}}

\def\r@ferr{\endreferences\errmessage{I was expecting to see
\noexpand\endreferences before now;  I have inserted it here.}}
\let\r@ferences=\references
\def\references{\r@ferences\def\endmode{\r@ferr\par\endgroup}}
\let\endr@ferences=\endreferences
\def\endreferences{\r@fcurr=0
  {\loop\ifnum\r@fcurr<\r@fcount
    \advance\r@fcurr by 1\relax\expandafter\r@fis\expandafter{\number\r@fcurr}%
    \csname r@ftext\number\r@fcurr\endcsname%
  \repeat}\gdef\r@ferr{}\endr@ferences}

\let\r@fend=\endpaper\gdef\endpaper{\ifr@ffile
\immediate\write16{Cross References written on []\jobname.REF.}\fi\r@fend}

\catcode`@=12

\def\(#1){(\call{#1})}
\def\ssc{\scriptscriptstyle}
\def\scc{\scriptstyle}
\def\-#1{_{\ssc {#1} }}
\def\s #1{{\cal {#1}}}

\def\hat #1{\mathaccent94{#1}}


\title INFLUENCE~FUNCTIONALS AND THE ACCELERATING~DETECTOR
\author J.R. Anglin\footnote{$^\dagger$}{anglin@hep.physics.mcgill.ca}
\affil Physics Department, McGill University
3600 University Street
Montreal, Quebec, CANADA H3A 2T8

\abstract The influence functional is derived for a massive scalar field in
the ground
state, coupled to a uniformly accelerating DeWitt monopole detector in $D+1$
dimensional Minkowski space.  This confirms the local
nature of the Unruh effect, and provides an exact solution to
the problem of the accelerating detector without invoking a non-standard
quantization.  A directional detector is presented which is efficiently
decohered by the scalar field vacuum, and which illustrates an important
difference between the quantum mechanics of inertial and non-inertial frames.
{}From the results of these calculations, some comments are made regarding the
possibility of establishing a quantum equivalence principle, so that the
Hawking
effect might be derived from the Unruh effect.

\endtitlepage

\eject

\head {Introduction}

A typical state of a quantum field, such as the vacuum, contains coherences
between spatially distant degrees of freedom.  This is the source of the
non-locality of measurement and other
non-intuitive results of quantum theory.  The ``decoherent histories"
reformulation of the theory allows
one to consider quantum mechanical systems which do not possess long range
coherence[\cite{dechis}], and other recent work likewise examines quantum
coherence specifically[\cite{coh, HPZ}].  Both of these areas of research
employ influence functionals as essential tools.
The influence functional for a scalar field initially in the
ground state, coupled to a point-like accelerating detector, is of interest
both because it provides another example of the use of influence functionals,
and because it sheds light on the acceleration-induced heating of the vacuum,
referred to as the Unruh effect[\cite{bell}].

The acceleration-induced heating of the vacuum was first discussed at length by
Unruh, as a toy model for Hawking radiation from an eternal black
hole[\cite{unruh}].
An earlier paper by Davies equipped flat space with a static, reflecting
boundary in order to model a black hole formed by collapse; this led to a
similar result[\cite{davies}].  The original derivation of this
effect begins by quantizing a scalar field in Rindler co-ordinates, replacing
ordinary time evolution with translation along trajectories of
constant proper acceleration[\cite{fulling, wald}].  From this point of view
the interaction between a point-like
accelerating detector and the field becomes a global problem: it involves
re-labelling the entire Fock space.  It also turns the accelerating observer's
personal event horizon into an apparently special location in Minkowski space,
leading to the impression that the thermal effects of acceleration are somehow
global properties of spacetime, rather than local effects.

On the other hand, arguments may be constructed that obtain the thermal
character of the vacuum, as seen by an accelerating detector,
without using the Rindler quantization[\cite{bell,
takagi1,jacobson}].  These discussions involve
time-dependent perturbation theory, and an examination of the two-point
function of a scalar field in terms of the detector's proper time.  They
suggest that the apparent horizon of the detector plays no
direct role in generating the Unruh effect.  This paper extends this line of
argument beyond
perturbation theory, deriving an exact result in the form of an influence
functional.  The physics involved is all implicit in the perturbative approach,
but the formalism used is more powerful and perhaps less familiar.

Proceeding from the idea that acceleration implies, in a local manner, an Unruh
temperature, it has been suggested that an appeal to the equivalence principle
might allow one to derive the Hawking effect from the Unruh
effect[\cite{jacobson}].
Influence functionals are also used in this paper to
investigate how useful a simple kind of quantum equivalence principle might be
as a basic tool, for possible application in curved spacetime.
The study of a particular model for a directional
accelerating detector shows that, in addition to the Unruh temperature and the
differences between Rindler and Minkowski densities of states, there
is another significant difference between the
local behaviours of a quantum field perceived by accelerating and inertial
observers: the correspondence between directions in space and orthogonal field
modes breaks down in the accelerating frame.
There are such significant additional effects of acceleration in
quantum field theory beyond the appearance of
a temperature, that using an equivalence principle will not
necessarily be as helpful as one might hope.
This and other facts suggest that
further study is needed if a quantum equivalence principle is to
be used to approach quantum field theory in curved spacetimes.

This paper is organized as follows.
A brief summary of the method of influence functionals is
presented in Section 1.  In Section 2 the Unruh effect in 1+1
dimensions is derived using
influence functionals, and the role of spatial regions causally disconnected
from the detector is made clear in Section 3.
Section 4 extends the results of Section 2 to $D+1$ dimensions.  Section 5 then
deals with a directional detector in $2+1$ dimensions.  Section 6 concludes,
with some comments on the extension of this analysis to the case of Hawking
radiation.

\head {1.  Review of Influence Functionals}

The theory of influence functionals was presented in 1963 by R.P.~Feynman and
F.L.~Vernon,~Jr.  For a full discussion of the technique, the
reader is referred to their original paper[\cite{vernon}].

In many quantum mechanical problems, at least implicitly, one
has an observed system coupled to an environment which is unobserved.  Both
system and environment are quantum mechanical; one assumes that the complete
Hilbert space may be spanned by a basis of direct product states of the form
$|\Psi_{env}\rangle |\psi_{sys}\rangle$.  The probability $P\-{FI} =
\vert\langle F|\hat U|I\rangle\vert^2$ of a
transition from an initial
state with a product wave function $\Psi\-I\psi\-i$ to a final state
$\Psi\-F \psi\-f$ is given by the path integral
$$\eqalign{
P\-{FI} =
\int\! {\s D}\!E{\s D}\!E'{\s D}\!S{\s D}\!S'\,
&\Psi\-I{\scc (E\-i)}\psi\-i{\scc (S\-i)}\Psi^*\-I{\scc (E'\-i)}\psi^*\-i
{\scc (S'\-i)}\Psi^*\-F{\scc (E\-f)}
\psi^*\-f{\scc (S\-f)}\Psi\-F{\scc (E'\-f)}\psi\-f{\scc (S'\-f)}\cr
&\times\  \exp {i\over\hbar}\!
\int_{t\-i}^{t\-f}\!dt\,\left( L(E,S) - L(E',S')\right)\;. }\eqno(PPI)
$$
Here $E$ and $E'$ variables represent the environment degrees of freedom, while
$S$
and $S'$ stand for the system under observation.  $E\-i$ stands for $E(t\-i)$,
and similarly for the other variables and subscript.  The Lagrangian is
$$
L(E,S) = L\-s(S) + L\-e(E) + L\-{int}(E,S)\;,
$$
organized into a system term, an environment term, and an
interaction term.

By axiom, the transition probabilities are to be summed
over all final states of the unobserved sector.  This leaves a truncated
quantum
theory describing only the observed degrees of freedom.  (The initial state of
the environment must be given, of course; it modifies the parameters of the
truncated theory\footnote{$^\dagger$}{It is assumed throughout this paper that
the system and the environment are initially uncorrelated, so that the initial
state of the universe may be written as a product of system and environment
states.  For systems strongly coupled to their environment, this assumption is
not realistic[\cite{HPZ}].  In the present case, a hypothetical detector will
be used to probe the unperturbed state of a field, and so the uncorrelated
initial state is appropriate.}.)  A great virtue of
Feynman's path
integral formulation of quantum mechanics, in comparison with the canonical
approach, is that one can in principle carry out this truncation
before considering the evolution of the observed sector.  By performing first
the path integral over all the unobserved degrees of freedom, one is left with
a modified path integral containing only observed degrees of freedom.  This
modified path integral is temporally non-local, and it provides non-unitary
time
evolution, since its construction has involved replacing pure final states
with  decoherent mixtures in which all states of the unobserved sector are
equally probable.

The part of the truncated transition probability path integral that contains
the non-local and non-unitary evolution is called the {\it influence
functional}
$F[S,S']$.  It
describes completely the influence of the unobserved environment on the
observed system.  One can write
$$
P\-{fi} = \int\!{\s D}\!S{\s D}\!S'\,\psi\-i{\scc (S\-i)}\psi^*\-i{\scc
(S'\-i)}
\psi^*\-f{\scc (S\-f)}\psi\-f{\scc (S'\-f)}\,
F[S,S']\,
e^{{i\over\hbar}\!\int_{t\-i}^{t\-f}\!dt\,\left(
L\-s(S) - L\-s(S')\right)}\;\; , \eqno(TPI)
$$
where $P\-{fi}$ denotes the probability of a transition of the observed
system.  Comparing \(TPI) and \(PPI), and using the fact that $\sum_F
|F\rangle\langle F|$ is an identity operator, it is straightforward to obtain
the formula for the influence functional:
$$\eqalign{
F[S,S']
 =& \int\!{\s D}\!E{\s D}\!E'\,\Psi\-I{\scc (E\-i)}\Psi^*\-I{\scc (E'\-i)}
\,\delta{\scc (E\-f-E'\-f)}\cr
&\;\;\times\exp {i\over\hbar} \!\int_{t\-i}^{t\-f}\!dt\,
\left( L\-e(E) - L\-e(E') + L\-{int}(E,S) - L\-{int}(E',S') \right) \cr
&= \langle \Psi\-I| \hat U^\dagger\-{S'}(t\-f,t\-i)
\hat U\-S(t\-f,t\-i)|\Psi\-I\rangle\; .}
\eqno(if)
$$
$\hat U\-S$ refers to the unitary time evolution operator for the environment,
with $S$ in $L\-{int}(E,S)$ treated as an external source.  (Because $S$ and
$S'$ are different, $\hat U\-{S'}^\dagger$ is not the inverse of $\hat U\-S$.)

The influence functional can also be used to obtain the time evolution of the
observed system density matrix $\rho(S,S';t)$:
$$
\rho(S\-f,S\-f';t\-f) = \int\!{\s D}\!S{\s D}\!S'\,\rho(S\-i,S\-i';t\-i)
F[S,S']\,
e^{{i\over\hbar}\!\int_{t\-i}^{t\-f}\!dt\,\left(
L\-s(S) - L\-s(S')\right)}\;.\eqno(dmev)
$$
Because $F[S,S']$ is a non-unitary kernel, it provides a mechanism for
decoherence in the evolution of $\rho$; one can interpret this effect as
representing the loss of information from the observed system into the
unobserved environment.  It is in this decohering context that influence
functionals have recently found application[\cite{dechis,coh}].

A pertinent example of an influence functional is that for a collection of
harmonic oscillators, distributed over all frequencies $\omega$ with spectral
density
$G(\omega)$, linearly coupled to an observed degree of freedom $Q(t)$.  The
Lagrangian for the unobserved sector, represented by the variables $X\-\omega$,
is of the form
$$
L = \int\!d\omega \,G(\omega)\,\int\!dt\left({1\over2}\dot X\-\omega^2 -
{1\over2}X\-\omega^2 + QX\-\omega\right)\;\; .\eqno(lho)
$$
{}From this Lagrangian, and choosing the initial state of the collection of
oscillators to be thermal, one may obtain the influence
functional[\cite{vernon}]
$$\eqalign{
\ln &F[Q,Q'] = -{1\over2\hbar}\int\!{d\omega\over\omega}
G(\omega)\int_{t\-i}^{t\-f}\!dt \int_{t\-i}^t\!dt' \left[
Q(t)-Q'(t)\right]\cr
&\times\ \left(\coth{\hbar\omega\over2kT}\left[Q(t')-Q'(t')\right]
\cos\omega(t-t')
-i\left[Q(t')+Q'(t')\right]\sin\omega(t-t')\right)\;,}\eqno(tif)
$$
where $T$ is the temperature of the initial state of the ensemble of
oscillators, and $k$ is the Boltzman constant.  Note that the
$t'$ integral is over the range $t'<t$.

Another useful instance of an influence functional is the one describing a
free massive scalar field in 1+1 dimensions, initially in the vacuum state,
linearly coupled to an observed degree of freedom $Q$.  The Lagrangian of
the scalar field may be written
$$
L = {1\over2\pi}\!\int_{-\infty}^{\infty}\!dk\,{1\over2}\left(
{1\over c^2}\dot\Phi\-k\dot\Phi^*\-k
- {\omega\-k^2\over c^2}\Phi\-k\Phi^*\-k +
Q(t)[A^*\-k(t)\Phi\-k+A\-k(t)\Phi^*\-k
]\right)\;\; .
\eqno(lsf)
$$
Here $\Phi\-k(t)$ is the $k$th Fourier mode of the field, and
$\omega\-k = c\sqrt{k^2+m^2}$ for $\hbar m / c$ the field mass.  $A\-k$ is a
coupling strength which may depend both on $k$ and on time $t$.
For this Lagrangian, one obtains the influence functional
$$\eqalign{
F[Q,Q'] &= \exp -{c\over4\pi\hbar}\int_{-\infty}^{\infty}\!
{dk\over\sqrt{k^2+m^2}}\ V\-k[Q,Q']\qquad\hbox{where}\cr
V\-k[Q,Q'] &= \int_{t\-i}^{t\-f}\!dt\int_{t\-i}^t\!dt'\ A\-k(t)A^*\-k(t')\,
\left[Q(t)-Q'(t)\right]\cr
&\;\times\ \left(\left[Q(t')-Q'(t')\right]
\cos\omega(t-t') -i\left[Q(t')+Q'(t')\right]\sin\omega(t-t')\right)\;.}
\eqno(sfif)
$$
Equations \(tif), \(lsf) and \(sfif) will all be used below.

\head{2. The Accelerating Detector in the Minkowski Vacuum}

A pointlike quantum mechanical detector may be idealized as interacting with
its
environment only through the DeWitt monopole moment $Q(t)[\cite{dewitt}]$.  If
it is linearly coupled to an otherwise
free massive scalar field $\phi$ in 1+1 dimensional Minkowski space, then it
serves
as a localized probe of the field.  If the detector is constrained to move
along a trajectory $x = x\-0(\tau)$, $t = t\-0(\tau)$, where $x$ and $t$ are
Cartesian co-ordinates and $\tau$ is the detector's proper time, then one has
the interaction Lagrangian
$$
L\-{int}(\phi, Q,t) = \int\!d\tau\ Q(\tau)\delta\bigl(t-t\-0(\tau)\bigr)
\int\!dk\,
\Phi\-k\bigl(t\-0(\tau)\bigr) e^{-ikx\-0(\tau)}\;\;.\eqno(lint1)
$$
The delta
function will clearly serve to transform inertial time integrals into proper
time integrals, in such formulas as \(tif) and \(sfif).  When this is done, it
can be seen that \(lint1) gives the simple action term
$$
{\s S}\-{int} = \int\!d\tau\, Q(\tau)\phi\bigl(x\-0(\tau),
t\-0(\tau)\bigr)\;.
$$

If the detector's trajectory is one of constant acceleration $a$, then
$$\eqalign{
x\-0(\tau) &= {c^2\over a}\cosh{a\tau\over c}\cr
t\-0(\tau) &= {c\over a}\sinh{a\tau\over c}\;\;.}\eqno(traj)
$$
Inserting this into \(lint1) and comparing with \(lsf), one obtains the
time-dependent, $k$-dependent coupling of the detector to the field:
$$
A\-k(\tau) = e^{i{kc^2\over a}\cosh{a\tau\over c}}\;\;.\eqno(ak)
$$
Applying this in turn in \(sfif), and using the delta functions to replace
$t$ and $t'$ everywhere with the functions of proper times $t\-0(\tau)$ and
$t\-0(\tau')$, one derives the
influence functional of the free massive scalar field on the pointlike
accelerating detector.
$$\eqalign{
F&[Q,Q'] =\cr &\exp
-{c\over2\hbar}\int_{\tau\-i}^{\tau\-f}\!d\tau\!\int_{\tau\-i}^{\tau}\!d\tau'
\,[Q(\tau)-Q'(\tau)]\,[Q(\tau')U(\tau,\tau') - Q'(\tau')U^*(\tau,\tau')]\; ,}
\eqno(if1)
$$
where
$$
U(\tau,\tau') \equiv {1\over2\pi}\!\int_{-\infty}^{\infty}\!{dk\over
\sqrt{k^2+m^2}}
e^{-i{kc^2\over a}[\cosh{a\tau\over c} -
\cosh{a\tau'\over c}] - i{\omega c\over a} [\sinh{a\tau\over c} -
\sinh{a\tau'\over c}]}\;\; . \eqno(u)
$$

Equation \(if1) is an implicit answer to the question of how the ground state
of the scalar field appears to an accelerating observer.  One need only
proceed with the path integral \(TPI), using some appropriate Lagrangian
$L\-s(Q)$ for
the detector itself, to have an exact solution for the detector's time
evolution.  Rather than doing this, however, it will be more instructive to
derive the Fourier transform of the apparently complicated function
$U(\tau,\tau')$.  The result will be of the same form as \(tif),
providing a clear interpretation of the influence functional \(if1) as
representing the effect of a heat bath at the Unruh temperature $kT = {\hbar
a\over2\pi c}$.

Redefining $k = m\sinh(\eta - a{\tau +\tau'\over2c})$, \(u) becomes
$$\eqalign{
U(\tau,\tau') &= {1\over2\pi}\!\int_{-\infty}^{\infty}\!d\eta\,
e^{-i{mc^2\over a}[\sinh(\eta +
a{\tau-\tau'\over2c}) - \sinh(\eta - a{\tau-\tau'\over2c})]}\cr
&={1\over2\pi}\!\int_{-\infty}^{\infty}\!d\eta \,e^{-2i{mc^2\over a}\cosh\eta\,
\sinh({a\over c}{\tau-\tau'\over2})}\;\; .}\eqno(u1)
$$
One now invokes a crucial identity from the theory of modified Bessel functions
of imaginary order\footnote{$^\dagger$}{These functions appear naturally in
solutions to the scalar wave equation in Rindler co-ordinates, and therefore in
wave functions for a scalar field quantized in an accelerating
frame[\cite{fulling}].} (denoted $K_{i\nu}$)[\cite{K1}]:
$$
e^{-i\alpha\sinh{x\over2}} = {4\over\pi}\int_0^\infty\!d\nu\,K_{2i\nu}(\alpha)
[\cosh(\pi\nu)\cos(\nu x) - i \sinh(\pi\nu)\sin(\nu x)]
\;\;.\eqno(K1)
$$
Substituting $\nu\to {c\omega\over a}$, this allows us to re-express \(if1) as
$$\eqalign{
&F[Q,Q'] = \exp -{1\over2\hbar}\int\!{d\omega\over\omega}\,
G(\omega)\int_{\tau\-i}^{\tau\-f}\!d\tau \int_{\tau\-i}^\tau\!d\tau' [Q(\tau)
-Q'(\tau)]\cr
&\times\ \left(\coth{c\pi\omega\over a}[Q(\tau')-Q'(\tau')]
\cos\omega(\tau-\tau')
-i[Q(\tau')+Q'(\tau')]\sin\omega(\tau-\tau')\right)\ ,}
\eqno(atif)
$$
where
$$\eqalign{
G(\omega)&\equiv {2c^2\omega\over a\pi^2}\int_{-\infty}^\infty\!d\eta
\,K_{2{ic\omega\over
a}}(2{\scc {mc^2\over a}}\cosh\eta) \sinh{c\pi\omega\over a}\cr
&= {2c^2\omega\over a\pi^2}\sinh{c\pi\omega\over a}\ [K_{ic\omega\over a}
({\scc {mc^2\over
a}})]^2\;\; ,}\eqno(Gnu)
$$
using another modified Bessel function identity[\cite{K2}].

Comparing \(atif) with \(tif), one observes that the effect of the scalar field
vacuum on an accelerating detector is exactly that of a heat bath at
temperature $kT = {\hbar a\over2\pi c}$.

In order to verify that these results coincide with the standard conclusion
that an accelerating observer sees the Minkowski vacuum as a thermal ensemble
of Rindler modes[\cite{unruh, wald}], it is worth determining what $G(\omega)$
would appear in a thermal
influence functional derived using Rindler quantization[\cite{fulling}].  Begin
with the action for a scalar field $\phi(\rho,\xi)$ in Rindler
co-ordinates[\cite{rindler}]
$$
{\s S} =  \int\!d\xi\,\left[\int_0^{\infty}\!{\rho d\rho\over2c}
\left(\rho^{-2}
\partial_\tau\phi^2
-\partial_\rho\phi^2 - m^2\phi^2 \right)\ + {\scc{c\over a}}Q({\scc{c\over a}}
\xi)\, \phi({\scc {c^2\over a}},\xi)\right]\;,
\eqno(sflag)
$$
where $\rho$ and $\xi$ are related to
the Cartesian co-ordinates used above by the definitions
$$\eqalign{x &= \rho\cosh\xi\cr
t &=\rho\sinh\xi\;.}\eqno(xirho)
$$
These co-ordinates label only the Rindler wedge
of Minkowski space $x>0$, $|t|<x$.  The Minkowski line element is given by
$$
ds^2 = d\rho^2 -\rho^2d\xi^2\;.
$$

Following Fulling[\cite{fulling}], one now expands the scalar field
$\phi$ in Rindler modes $\varphi\-\nu(\xi)$ instead of Fourier modes
$\Phi\-k(t)$.  The $\varphi\-\nu$ are the co-efficients in an expansion of the
field $\phi$ in solutions of the scalar wave equation
$$
\rho^2{d^2\phi\over d\rho^2} + \rho{d\phi\over d\rho} - m^2\rho^2\phi =
{d^2\phi\over d\xi^2} = -\nu^2\phi
$$
separated in the $\rho,\xi$ variables.  As mentioned above,
these solutions are modified Bessel functions of
imaginary order, and so the Rindler modes are defined by
$$
\phi(\rho,\xi) = {1\over\pi}\int_0^\infty\!d\nu \sqrt{2\nu\sinh\pi\nu}
K_{i\nu}(m\rho) \varphi\-\nu(\xi)\;\; . \eqno(phinu)
$$
One can then use the orthogonality relation[\cite{fulling}]
$$
{1\over\pi^2}\int_0^\infty\!{d\rho\over\rho} K_{i\mu}(m\rho)\,K_{i\nu}(m\rho)
=  {\delta(\mu -\nu)\over2\nu\sinh\pi\nu} \eqno(korthog)
$$
and the modified Bessel equation satisfied by $K_{i\nu}(m\rho)$ to re-write
\(sflag) as
$$
{\s S} = \int\!{d\xi\over c}\int_0^\infty\!d\nu\,\left[{1\over2}
\partial_\xi\varphi\-\nu^2 -
{1\over2}\nu^2\varphi\-\nu^2 + {c^2\over a\pi} Q({\scc{c\over a}}\xi)
\sqrt{2\nu\sinh\pi\nu}K_{i\nu}({\scc {mc^2\over a}}) \varphi\-\nu\right]\;\; .
\eqno(rmodlag)
$$

The fastest way to obtain the influence functional for a thermal ensemble of
Rindler modes with this action is to put \(rmodlag) into the same form as the
action derived from \(lho).  This requires that the time
variable be the proper time of the $Q$ system, and that the interaction term
have the same weight as the kinetic term.  Both conditions may be arranged by
re-scaling $\xi ={a\over c}\tau$, $\varphi\-\nu = {a\over\pi}
X\-\nu\sqrt{2\nu\sinh\pi\nu}K_{i\nu}({mc^2\over a})$,
and $\nu = {c\over a}\omega$.  One obtains ${\s S}=\int\!d\tau\,L(\tau)$, where
$$
L(\tau) = \int_0^\infty\!d\omega\, {2c^2\omega\over a\pi^2}
\sinh{c\pi\omega\over a}
\,[K_{ic\omega\over a}({\scc
{mc^2\over a}})]^2\, \Bigl[{1\over2}\dot X\-{\omega}^2 -
{1\over2}\omega^2X\-\omega^2 + Q X\-\omega\Bigr]\;\; . \eqno(rho)
$$
The measure on $\omega$ appearing here is exactly $G(\omega)$ from \(Gnu).
Comparison with \(lho), \(tif) and \(atif) then shows that a heat bath of
Rindler modes
at the Unruh temperature indeed gives the same influence functional as was
derived using ordinary Fourier modes populated at zero temperature.

\head{3.  The Role of Long-Range Coherence}

This section examines the role played in generating the Unruh effect by those
of the scalar field's degrees of freedom that lie outside the past light cone
of the trajectory of the detector; this is equivalent to considering the role
of vacuum state coherences between degrees of freedom inside and outside the
light cone. The past light cone of a constant-acceleration trajectory from
$\tau = -\infty$ to $\tau = \infty$, with the past horizon as an initial data
surface, of course contains the whole Rindler wedge.  If one wishes to consider
the evolution of the detector over all time, then the formulation of the Unruh
problem in terms of  Rindler modes seems correctly to attach importance to the
horizon.  But \(atif) describes a thermal environment, regardless of $\tau\-i$
and $\tau\-f$: an
accelerating detector feels a heat bath over any time interval, however
short\footnote{$^\dagger$}{The time necessary for the detector to reach
equilibrium with the environment is a separate problem, not considered here.}.
Hence one can consider the behaviour of the detector between any two finite
proper times $\tau\-i$ and $\tau\-f$, and the Unruh effect will still appear.
The past light cone of a finite trajectory does not coincide with the Rindler
horizon; consequently the role of quantum coherences between field degrees of
freedom separated by space-like intervals is distinct from the role of the
horizon.   The former may best be studied using the influence functional
formalism.

To begin, consider the simple case of a harmonic oscillator driven by an
external force $J(t)$, with the Lagrangian
$$
L_{HO}={1\over2} \dot q^2 - {1\over2}\Omega^2 q^2 + Jq \; .\eqno(qlag)
$$
The amplitude for a transition between initial and final $q$-states is
$$\eqalign{
&\langle q\-f|\hat U_J(t\-f,t\-i) |q\-i\rangle \cr
=& \exp
{i\over\hbar}{1\over\sin\Omega(t\-f-t\-i)}
\Bigl[(q\-f^2+q\-i^2)\Omega\cos\Omega (t\-f-t\-i) - 2q\-i q\-f\cr
&+2\int_{t\-i}^{t\-f}\!dt'\,J(t')[q\-f \sin\Omega t' +q\-i\sin\Omega(t\-f-t')]+
 \;\hbox{...}\;\Bigr]\; ,}\eqno(qamp)
$$
where the ellipsis denotes terms independent of $q\-i$ and $q\-f$.

The influence functional for the oscillator, on the system represented by $J$,
is given by
$$
F[J,J'] = \int\!dq\-i dq\-i'\,\psi\-i(q\-i)\psi\-i^*(q\-i') \int\!dq\-f\,
\langle q\-i'|\hat U_{J'}^\dagger |q\-f\rangle\langle q\-f|\hat U_J
|q\-i\rangle\;,\eqno(kerker)
$$
where $\psi\-i$ is the wave function for the initial state of the oscillator.
One may extract and simplify the kernel in this definition, writing
$$\eqalign{
&\int\!dq\-f\,\langle q\-i'|\hat U_{J'}^\dagger |q\-f\rangle
\langle q\-f|\hat U_J
|q\-i\rangle\cr
&={\s A}[J,J']\,
\delta\Bigl(q\-i-q\-i'-{1\over\omega}\int_{t\-i}^{t\-f}\!dt\,
[J(t)-J'(t)]\sin\omega(t-t\-i)\Bigr)\cr
&\qquad\qquad\times \exp
{i\over\hbar}{q\-i+q\-i'\over2}\int_{t\-i}^{t\-f}\!dt\,
[J(t)-J'(t)]\cos\omega(t-t\-i)\;.}\eqno(kerker2)
$$
Since $q\-f$ is the same in both amplitudes in the first line, the $q\-f^2$
term in \(qamp) is cancelled by the hermitian conjugate term, leaving an
exponent linear in $q\-f$.  The integration over final states then converts
this into a delta function, which is then used to eliminate $(q\-i-q\-i')$ from
the expression.  The prefactor ${\s A}$ is independent of both $q\-i$ and
$q\-i'$.

This analysis may be generalized to the case of the massive scalar field,
coupled as prescribed by \(lsf), \(lint1), and \(ak), since
the Fourier modes $\Phi\-i(k)\equiv\Phi\-k(t\-i)$ are closely analogous to the
Schroedinger picture position operators of an infinite
set of decoupled oscillators.  One obtains for the kernel analogous to
\(kerker2)
$$\eqalign{
&\langle\Phi\-i'(k)|\hat U^{\ssc\dagger}\-{Q'}\hat U\-Q|\Phi\-i(k)\rangle
    ={\s A}[Q(\tau),Q'(\tau)]\cr
&\ \times \prod_{k}\delta\Bigl(\Phi\-i(k)-\Phi\-i'(k) -
\!\int_{\tau\-i}^{\tau\-f}\!d\tau\,
{Q-Q'\over\sqrt{2\pi}\sqrt{k^2+m^2}}
e^{ikx\-0(\tau)}\sin\omega\-k[t\-0(\tau)-t\-i]\Bigr)\cr
&\ \times\exp{ic\over\sqrt{2\pi}\hbar}\int_{-\infty}^\infty\!dk{\Phi\-i(k)
+\Phi\-i'(k)\over2}\int_{\tau\-i}^{\tau\-f}\!d\tau\,(Q-Q')
e^{ikx\-0(\tau)}\cos\omega\-k[t\-0(\tau)-t\-i]\;.}\eqno(kerphi)
$$
Fourier transforming back to field variables, so that $\Phi\-i(k)$ is replaced
by $\phi\-i(x)$, for $x$ the inertial spatial co-ordinate mapping the initial
data surface $t=t\-i$, this amplitude may be re-written
$$\eqalign{
&\langle\phi\-i'(x)|\hat U^{\ssc\dagger}\-{Q'}\hat U\-Q|\phi\-i(x)\rangle
={\s A}[Q(\tau),Q'(\tau)]\cr
&\ \times\prod_{x}\delta\Bigl(\phi\-i(x)-\phi\-i'(x)-
\int_{\tau\-i}^{\tau\-f}\!d\tau\,(Q-Q')
G[x\-0(\tau-x,t\-0(\tau)-t\-i]\Bigr)\cr
&\ \times\exp
-{i\over\hbar}\int_{-\infty}^\infty\!dx\,{\phi\-i(x)+\phi\-i'(x)\over2}
\int_{\tau\-i}^{\tau\-f}\!d\tau\,(Q-Q'){\partial\over\partial t\-i}
G[x\-0(\tau)-x,t\-0(\tau)-t\-i]\;,}\eqno(kerchief)
$$
where the odd Green's function for the scalar wave equation is defined by
$$
G[x,y] \equiv
{1\over\sqrt{2\pi}}\int_{-\infty}^\infty\!{dk\over\sqrt{k^2+m^2}}\,e^{ikx}
\sin\omega\-k t\;.\eqno(gret)
$$
$G[x,t]$ vanishes outside the light cone $|ct|>|x|$.

Define the intersection of the light cone of the accelerating
trajectory with the initial data surface to be
the initial data region $S$.  (See Figure~1, where $t\-i=0$ is
chosen for convenience.)  The fact that $G$ vanishes outside the light cone
may then be used to re-write \(kerchief) as
$$
\langle\phi\-i'(x)|\hat U^{\ssc\dagger}\-{Q'}\hat U\-Q|\phi\-i(x)\rangle
= {\s K}\-S[\phi,\phi']\times \prod_{x\notin
S}\delta\Bigl(\phi\-i(x)-\phi\-i'(x)\Bigr)\;,\eqno(kerlast)
$$
where ${\s K}\-S$ is a functional of the field degrees of freedom lying inside
the initial data region $S$.

The influence functional may formally be obtained from \(kerlast) by inserting
the initial state wave functionals $\Psi\-i[\phi\-i(x)]$ and
$\Psi\-i^*[\phi\-i'(x)]$, and then integrating:
$$\eqalign{
F[Q,Q'] &= \int \prod_{x}d\phi\-i(x)d\phi\-i'(x)
\Psi\-i[\phi\-i(x)] \Psi\-i^*[\phi\-i'(x)]
\langle\phi\-i'(x)|\hat U^{\ssc\dagger}\-{Q'}\hat
U\-Q|\phi\-i(x)\rangle\cr
&=\int\!\prod_{x\in S} d\phi\-i(x) d\phi_i'(x)\, {\s K}\-S[\phi,\phi']\cr
&\qquad\times
\int\!\prod_{x\notin S} d\phi\-i(x) d\phi_i'(x)\,
\delta(\phi\-i-\phi\-i') \Psi_i[\phi\-i(x)] \Psi_i^*[\phi\-i'(x)]\cr
&\equiv\int\!\prod_{x\in S} d\phi\-i(x) d\phi_i'(x)\, {\s K}\-S[\phi,\phi']
\rho\-S[\phi\-i(x),\phi\-i'(x)]\;,}\eqno(fqqred)
$$
where the reduced density matrix $\rho\-S$ describing the state of the field
degrees of freedom lying within the region $S$ is defined implicitly in the
last line. The product over the continuous index $x$ is meant to describe  the
infinite-dimensional integrals over all initial field configurations
$\phi\-i(x)$ and $\phi\-i'(x)$, using the ordinary measure $d\phi\-id\phi\-i'$
in order to avoid confusion with path integrals over $\phi(x,t)$.

In general, initial data for a scalar field influence functional is provided by
the density matrix for the field on  the initial data surface, which in this
case is the $x$-axis.   Where the field is initially in a pure state $\Psi\-i$,
one has the initial density matrix $\rho\-i(\phi,\phi') =
\Psi\-i[\phi]\Psi^*\-i[\phi']$.  In the particular case of a point-like
detector, however, \(fqqred) shows that the influence functional involves no
more than a trace over the initial values of those  field degrees of freedom
lying outside the past light cone of the detector.  In this case, therefore,
the reduced density matrix $\rho\-S$, formed by taking the trace of $\rho\-i$
over all the degrees of freedom outside $S$, provides sufficient initial data
for the evolution of the detector between the given initial and final times.

Of course $S$ may be extended outside the past light cone  as far as one likes,
and it will still be true that the influence functional involves only a trace
over the field modes outside $S$.  Defining $\rho\-i$,  and then obtaining
$\rho\-S$ by  tracing out the modes of the field which lie outside the initial
data region,  is therefore a procedure which makes no reference whatever to the
detector's  trajectory or event horizon, and involves only ordinary, inertial
quantization. Thus, even though the long-range coherences of the scalar field
ground state  do affect the form of the reduced density matrix, it can
nevertheless clearly be seen that the  Unruh effect does not arise from any
interaction between the long-range coherence of the vacuum and the geometry of
the Rindler wedge.  On the contrary, the Unruh effect originates in the
identity \(K1) relating functions of inertial and accelerating time.

\head{4. Extension to $D>1$ Spatial Dimensions}

The extension of the analysis of Section 2 to higher spatial dimensions is easy
to perform, and it provides an example which helps clarify a confusing
difference  between the heat bath of Rindler modes and ordinary thermal
radiation.

The accelerating detector in $D>1$ spatial dimensions may be described by
amending \(u) so that
$$\eqalign{
{dk\over 2\pi}&\to {d^Dk\over (2\pi)^D}\cr
\sqrt{k^2+m^2}&\to \sqrt{|\overrightarrow k|^2+ m^2}\;.}
$$
By expressing the (D--1)-dimensional vector $(k\-2,k\-3,...,k\-D)$ in polar
form, so that
$$\eqalign{
k\-2 &\equiv r\cos\alpha\-1\cr
k\-3 &\equiv r\sin\alpha\-1\cos\alpha\-2\;,\;\;\hbox{etc.,}}
$$
one obtains for $U(\tau,\tau')$ an integral over $k\-1$, $r$, and the angles
$\alpha\-n$.  The angular integration is trivial, and contributes only a factor
equal to the area of the unit (D--2)-sphere.  The $k\-1$ integral may be
treated exactly as the $k$ integral in Section 3, except that $r$ modifies the
mass term by changing $m\to\sqrt{m^2+r^2}$.  One therefore obtains once again
$$\eqalign{
F\-D[&Q,Q'] =\cr &\exp -{1\over2\hbar}
\int_{\tau\-i}^{\tau\-f}\!d\tau \int_{\tau\-i}^\tau\!d\tau'\, [Q(\tau)
-Q'(\tau)]\,[Q(\tau')U\-D(\tau,\tau') - Q'(\tau')U\-D^*(\tau,\tau')]
}\eqno(Datif)
$$
and
$$
U\-D(\tau,\tau') =
\int\!{d\omega\over\omega}\,G\-D(\omega)
\left(\coth{c\pi\omega\over a}
\cos\omega(\tau-\tau') -i \sin\omega(\tau-\tau')\right)\;.\eqno(ud)
$$
For $D>1$, however, one now has
$$
G\-D(\omega) = {c^2\omega\sinh{c\pi\omega\over a} \over 2^{D-3}a\pi^{D+3\over2}
\Gamma({\scc{D-1\over2}})}
\int_0^\infty\!dr\, r^{D-2}[K_{ic\omega\over a}({\scc {c^2\over
a}}\sqrt{m^2 + r^2})]^2\;.\eqno(GDnu)
$$

For the case $m=0$ the $r$ integral may be obtained in closed form.  This case
is analyzed in \reff{takagi1}, where a response function related to
$U(\tau,\tau')$ is derived which is proportional,  for odd $D$, to the Planck
distribution function.  For even $D$, however, the  Planck factor is replaced
by the Fermi-Dirac distribution function, even though the problem involves a
bosonic field.  It is instructive to see how this comes about, for the massless
scalar field does not in fact seem fermionic to an accelerating observer in any
number of spatial dimensions.

The equations needed to evaluate \(GDnu) in the massless limit are[\cite{K3}]
$$\eqalign{
\int_0^\infty\!dr\, r^{D-2}[K_{i\nu}(\lambda r)]^2 &= {2^{D-4}\lambda^{1-D}
\over\Gamma(D-1)}
|\Gamma({\scc{D-1\over2}}+i\nu)|^2 [\Gamma({\scc{D-1\over2}})]^2\cr
  &={\pi^{3\over2}\lambda^{1-D}\over e^{\pi\nu} \pm e^{-\pi\nu}}
{\Gamma({\scc{D-1\over2}})\over2\Gamma({\scc{D\over2}})} P\-D(\nu) \; ,}
\eqno(K3)
$$
where the plus (minus) sign applies for $D$ even (odd).  $P\-D(\nu)$ is the
polynomial
$$\eqalign{
P_2 =&\  1\cr
P_{2d}(\nu) =& \ \prod_{n=1}^{d-1}[(n-{1\over2})^2+\nu^2],\;\;d>1\cr
P_{2d+1}(\nu) =&\ {1\over\nu}\prod_{n=0}^{d-1}[n^2+\nu^2]\;.}\eqno(Pnu)
$$
{}From these facts one obtains
$$\eqalign{
G\-D(\omega)\left.\right\vert_{m=0} &= {\omega P\-D({\scc{c\over a}}\omega )
\over 2^{D-1}
\pi^{D\over2}\Gamma({\scc{D\over2}})}
\left({a\over c^2}\right)^{D-2}
  {e^{{2\pi c\over a}\omega}-1\over
e^{{2\pi c\over a}\omega} + (-1)^D}\cr
&\equiv {\omega R\-D(\omega)\over\pi} {e^{{2\pi c\over a}\omega}-1\over
e^{{2\pi c\over a}\omega} + (-1)^D} \;,}\eqno(Gm0)
$$
where $R\-D$ is simply a re-scaled version of the polynomial $P\-D$.

The response function $F\-D(\omega)$ (equal to $F\-n$ of \reff{takagi1} for
$n=D+1$) is defined as
$$\eqalign{
F\-D(\omega) &= \lim_{L\to\infty} {1\over2L}\int_{-L}^L\!d\tau\!\int_{-L}^L\!
d\tau'\, e^{-i\omega(\tau-\tau')} U(\tau, \tau') \cr
&={\pi\over2\omega} G\-D(\omega) \times {2\over e^{{2\pi c\over a}\omega} -1}
\;.}
\eqno(response)
$$                                                                          It
is therefore easy to see that, for $m=0$, the response function is equal to a
polynomial multiplied by either a Planck or a Fermi-Dirac factor,
$$
F\-D(\omega) = {R\-D(\omega)\over
e^{{2\pi c\over a}\omega} + (-1)^D}\;,\eqno(FD)
$$
in agreement with \reff{takagi1}.

The important point to realize in interpreting \(FD) is that if one ignores the
parity of $R\-D(\omega)$, then information about the real and imaginary parts
of $U(\tau,\tau')$ has been discarded.  Yet these real and imaginary parts will
play drastically different roles in the detector path  integral ({\it i.e.}
dissipation versus fluctuation), and so are  physically distinguishable. The
true signature of a thermal environment is found not by simply extracting the
most obvious statistical factor from the response function, but by taking the
ratio of its  even and odd parts.  As implied by \(tif), this ratio will be
$\coth(\hbar\omega /2kT)$ for a bosonic heat bath at temperature $T$.  Since
$R\-D$ has the same parity as $D$, \(FD) actually describes such a bath:
$$
{F\-D(-\omega) + F\-D(\omega)\over F\-D(-\omega) - F\-D(\omega)} =
\coth{\pi c\over a}\omega\; .\eqno(proof)
$$

This example illustrates that the thermal factor $\coth(\hbar\omega /2kT)$,
which appears only in the real part of $U\-D(\tau,\tau')$, and which reflects
the fact that each mode of the scalar field is thermally populated, ought not
to be confused with the factor $G\-D(\omega)$  multiplying both real and
imaginary terms, which describes the density of modes with different
frequencies\footnote{$^\dagger$}{The  distinction between the spectral density
and the thermal population was explained in Reference [\cite{unruh2}]; the
influence  functional formalism further shows that the distinction is clearly
expressed in the phase of $U(\tau,\tau')$.}.  $G\-D(\omega)$ is such that the
scalar field heat bath felt by an accelerating observer has  in general a
different energy spectrum from thermal radiation in an inertial frame.
Nevertheless,  the scalar field does appear as a bath of bosonic oscillators
populated thermally at the Unruh temperature, and this is true for all values
of $D$.

The next section illustrates another, and perhaps more profound, difference
between the acceleration heat bath and inertial thermal radiation.

\head{5. A Directional Detector Exhibiting Decoherence}

The accelerating detectors studied so far have been coupled to the field
omni-directionally: they do not discriminate between different directions in
space.  In the process of determining that the acceleration heat bath is
anisotropic, various authors have considered directional detectors, whose
coupling singles out preferred  angles[\cite{HPD,israel,takagi2}].   This
section examines a model slightly different from previous ones.  This model has
a novel feature that is shown clearly in the influence functional  formalism:
if the detector's narrow line of sight is pointed away from its direction of
acceleration, then the influence of the scalar field vacuum  on the detector is
not like that of a stationary ensemble of oscillators.   Instead, the field
acts as a quantum measurement device that causes the detector's quantum state
to evolve rapidly into a decoherent mixture of eigenstates of the monopole
moment operator $Q$.  In the limit where the  angular aperture of the detector
is extremely narrow, this is all  that the scalar field does.

The present directional detector is similar to that of \reff{takagi2} in that
its window of angular sensitivity is infinitesimal.  The latter model uses
Rindler modes to describe the scalar field, however.  Since these modes do not
possess a momentum quantum number in the direction of acceleration, the
preferred angle of Reference~[\cite{takagi2}] is defined by a ratio of
transverse momentum to energy.  Such an angle is not actually a direction in
space, even in the detector's rest frame.   In this respect the present model
is more like that of  \reff{israel}, in that it  describes a directional
detector in terms of its coupling to Fourier modes in an instantaneously
co-moving inertial frame\footnote{$^\dagger$}{The spatial as well as the
temporal frequencies perceived by the accelerating observer  differ from those
in the instantaneously co-moving inertial frame[\cite{moreau}]. The description
of the directional detector in terms of its coupling to Fourier modes in
instantaneously co-moving frames presumes that the detector performs its
angular discrimination within a sufficiently short distance for this effect to
be negligible.}.  The angle $\theta$ referred to in the discussion below will
therefore be a true spatial direction, held constant in the frame of the
accelerating detector.

Since the problem of the accelerating directional detector has azimuthal
symmetry, all of its significant features are encompassed  in the case $D=2$.
In two spatial dimensions, then, consider the coupling
$$
L\-{int}(\tau) = {Q(\tau)\over8\pi^2}\int\!{d^2k\over\sqrt{k\-1^2+k\-2^2+m^2}}
W(k\-1,k\-2;\theta,\tau) \Phi\-{k\-1,k\-2}(t=0)
e^{ik\-1x\-0(\tau) -i\omega t\-0(\tau)}\;.\eqno(coup)
$$
$W(k\-1,k\-2;\theta,\tau)$ is a projection operator that vanishes outside a
narrow window, of infinitesimal width $\epsilon$, around the polar angle
$\theta$. In the rest frame of the detector at proper time $\tau$, it is given
by
$$
W(k^\tau\-1,k^\tau\-2;\theta,\tau) = \int_0^\infty\!
x\,dx\!\int_{\theta}^{\theta+\epsilon}\!d\alpha\,
\delta(k\-1 - x\cos\alpha)\delta(k\-2 - x\sin\alpha)\;.\eqno(w1)
$$

The cases $\theta=0,\pi$ turn out to be difficult to evaluate in closed form,
and to involve no significant new phenomena.  It will therefore be assumed that
$\sin\theta >> \epsilon > 0$.

To calculate the influence functional, it will be necessary to have $W$ at two
different times $\tau$ and $\tau'$ expressed in the same inertial frame, since
the kernel $U(\tau,\tau')$ will become
$$\eqalign{
U(\tau,\tau';\theta) &= {1\over4\pi^2}\int\!{d^2k\over\sqrt{k\-1^2+k\-2^2+m^2}}
W(k\-1,k\-2;\theta,\tau) W(k\-1,k\-2;\theta,\tau')\cr
&\;\times\exp i\left(k\-1[x\-0(\tau)-x\-0(\tau')] - \omega(k\-1,k\-2)
[t\-0(\tau)-t\-0(\tau')]\right)\;.}\eqno(fth0)
$$
It will be convenient to choose for this frame the one in which the detector is
at rest at proper time $\bar\tau \equiv {1\over2}(\tau+\tau')$.
In this frame, the detector's locations at times $\tau$ and $\tau'$ are given
by
$$\eqalign{
x\-0(\tau) =  x\-0(\tau') =& {c^2\over a}\cosh\Delta\tau\cr
t\-0(\tau) = -t\-0(\tau') =& {c\over a}\sinh\Delta\tau\;,}\eqno(xt)
$$
for $\Delta\tau={1\over2}(\tau-\tau')$.  Boosting $(k\-1,k\-2)$ into the
$\bar\tau$ frame, one obtains
$$\eqalign{
W(k\-1,k\-2;\theta,\tau) &= \int_0^\infty
x\,dx\!\int_\theta^{\theta+\epsilon}\!d\alpha\,\delta(k\-2 - x\sin\alpha)\cr
&\;\times
\delta(k\-1\cosh\Delta\tau + \sqrt{x^2 + m^2}\sinh\Delta\tau
 - x\cos\alpha)\;.}\eqno(w2)
$$
$W(k\-1,k\-2;\theta,\tau')$ is given by a similar equation, in which the plus
sign in the argument of the last delta function is replaced by a minus sign.

To leading order in $\epsilon$, the narrow limits of integration on $\alpha$
in \(w2) simply impose $\alpha = \theta$.  Therefore, to
leading order in $\epsilon$, and for $\sin\theta>>\epsilon>0$, one can obtain
$$\eqalign{
W(k\-1,k\-2;\theta,\tau) &W(k\-1,k\-2;\theta,\tau')\simeq \cr
&{\epsilon^2\over\sin\theta}\int_0^\infty\!dx\,x^2
\delta(k\-2 - x\sin\theta)\delta(2\sqrt{x^2+m^2}\sinh\Delta\tau)\cr
 &\;\times\delta\bigl(k\-1\cosh\Delta\tau + \sqrt{x^2 + m^2}\sinh\Delta\tau
   - x\cos\theta\bigr)\cr
 &=\delta(\tau-\tau')
{\epsilon^2\over\sin\theta}\int_0^\infty\!{x^2\,dx\over\sqrt{x^2+m^2}}
\delta(k\-1 - x\cos\theta)\delta(k\-2 - x\sin\theta)\;.}\eqno(deltau)
$$

Substituting this and \(xt) into \(fth0), one obtains
$$\eqalign{
U(\tau-\tau';\theta) &={\epsilon^2\over4\pi^2\sin\theta}
\delta(\tau-\tau')\int_0^\infty\!{x^2\,dx\over x^2+m^2}\cr
&\equiv {\epsilon^2A\over\sin\theta}\delta(\tau-\tau')\;.}\eqno(fth1)
$$
$A$ is divergent as written, but it may be regulated in the manner
described in the Appendix, by assuming that the field-detector coupling has a
cut-off at some large Rindler energy.
The regulated version of $A$ is finite and positive.

The reason for the delta function on $(\tau-\tau')$ is that a polar angle such
as $\theta$ is not Lorentz invariant.  As viewed in the lab frame, a fixed
polar angle in the accelerating frame decreases with time (if it is between 0
and $\pi$), converging towards the direction of  acceleration[\cite{SRtext}].
Hence, if the detector's window is bounded by two very close angles, held
constant in the accelerating frame,  then this  window will not overlap itself
at different times.  Consequently the product of the two window projection
operators at different times vanishes  unless the two times are the same (to
zeroth order in $\epsilon$).

To leading order in $\epsilon$, the influence functional implied by \(fth1) is
simply
$$
F[Q,Q';\theta] = \exp - {\epsilon^2 A\over2\hbar\sin\theta}
\int_{\tau\-i}^{\tau\-f}\!d\tau\, [Q(\tau)-Q'(\tau)]^2
\;,\eqno(Fdec)
$$
When $F[Q,Q';\theta]$ is used to obtain the time evolution of the detector
density matrix $\rho(Q,Q')$, it clearly suppresses all paths
$[Q(\tau),Q'(\tau)]$ except those in which $(Q-Q')$ rapidly approaches zero.
The density matrix $\rho(Q,Q')$ evolves towards being diagonal: the influence
functional \(Fdec) drives the detector's quantum state towards a decoherent
mixture of eigenstates of the monopole moment operator $Q$.  Such decoherence
is a generic effect of  influence functionals for large environments acting on
small systems, but \(Fdec) is remarkable for its simplicity.  It decoheres the
detector in the $Q$-state  basis; it does so in an obvious and direct manner;
and it does nothing else.

The reason for this simple behaviour lies in the lab-frame time-dependence of
the detector's time-independent angular window.  The apparent variation in an
inertial frame of the angle of receptivity means that an accelerating detector
with a narrow aperture, directed away from the axis of motion, couples at each
instant to a succession of different (and orthogonal)  modes of the scalar
field.  This is the physical interpretation of \(Fdec).  It implies that the
association of a fixed direction in space with a given field mode, which is a
basic feature of quantum field theory in inertial frames, and is an important
element in our notion of quantum fields as representing particles breaks down
in non-inertial frames.

It can also be shown that the forward and backward ($\theta = 0$ or $\pi$)
directions, which represent orthogonal field modes in inertial quantization,
are associated with the same mode of the field in the Rindler quantization. So
not only can a fixed angle correspond to a succession of modes, but distinct
angles can correspond to a single mode.

The fact that the delta function in \(fth1) may be related to the elementary
relationship between inertial and accelerating angular directions should
confirm that it is not an unrealistic artifact of some approximation that has
been used.  If the calculations are extended to include terms of third order in
the narrow angular width $\epsilon$, then the delta function is indeed blurred
into a narrow window, but the basic features of \(Fdec) are not significantly
changed.  A brief sketch of the extension of \(fth1) and \(Fdec) to third order
follows.

One can obtain by straightforward means the result that
$$\eqalign{
U(\tau,\tau';\theta) =& {\epsilon^3\over\sin\theta} [\tilde
A\cot\theta  + iB\csc\theta]\delta(\tau-\tau')\cr
&+{\epsilon\over 4\pi^2\sin\theta}\int_0^\epsilon\!d\gamma\int_0^\infty\!dx\,
{x^2\over x^2+m^2} \delta\Bigl(\tau-\tau' - {x\sin\theta\over\sqrt{x^2+m^2}}
\gamma\Bigr) + {\s O}(\epsilon^4)\;,}\eqno(u3)
$$
where $\tilde A$ are $B$ are real constants defined by integrals needing
to be regulated in the same manner as that in \(fth1).  The imaginary term
proportional to $B$ and the smeared delta function lead to new terms in the
influence functional:
$$\eqalign{
F[Q,Q';\theta] &
= \exp
-{\epsilon^2\over2\hbar\sin\theta}\int_{\tau\-i}^{\tau\-f}\!d\tau\,\Bigl(
[A + \epsilon\cot\theta\tilde A][Q(\tau)-Q'(\tau)]^2\cr
&\; \qquad +i\epsilon
\csc\theta B[Q(\tau)^2-Q'(\tau)^2] + \epsilon C
{d\over d\tau}[Q(\tau)-Q'(\tau)]^2\Bigr)\;,}\eqno(fqqtheta3)
$$
where a new real constant $C$ is defined by the $x$ integral in \(u3).
The term in $F[Q,Q';\theta]$ proportional to $B$ is simply an addition to the
quadratic potential term in the $Q$-system Lagrangian $L(Q)$.  The term arising
from the smearing of the delta function also tends to decohere the system in
the $Q$ basis.

\head{6.  Conclusion}

The apparent heating of the vacuum observed along an accelerating trajectory
seems to be a fundamental result in non-inertial, relativistic quantum field
theory.  An acceleration provides a natural energy scale, but that this scale
should appear precisely as a temperature is an example of rare natural
simplicity.  The underlying reason for this simplicity seems originally to have
been sought in the behaviour of certain globally preferred co-ordinate systems
near event horizons[\cite{unruh,wald}]. This understanding of the phenomenon of
acceleration temperature had distressing implications, however, in that the
results of local measurements  seemed to depend on global properties of
spacetime.

The analysis presented above confirms that global issues  arise only when they
are explicitly invoked, in trying to discuss the state of a quantum field over
the entire Rindler wedge.  The thermal effects may be isolated from the global
problem by the use of influence functional methods, which allow one to
determine the properties of the field as probed by a point-like detector.  The
ultimate source of the acceleration temperature is then seen to be the relation
between sines and cosines of the proper times of inertial and accelerating
observers.  The Unruh effect may be considered a distant cousin to the inertial
forces of Newtonian mechanics, in that it is a property of the acceleration
itself.

The calculations that lead to this result require a reduced density matrix for
the initial state of the quantum field within a region of space.  In Minkowski
space this information is available because the correct ground state is known.
If these methods are to be extended to problems in curved spacetime, in
particular the Hawking radiation of black holes, some proposal for the field
ground state must be assumed.  As suggested by Jacobson[\cite{jacobson}], the
ground state might be constrained to possess certain properties measured by
local observers; for example, a static detector ``near" the horizon might be
required to behave as an accelerating detector in flat space.  This naive
quantum equivalence principle would provide the standard Hawking effect by
means of the Unruh effect.

One might hope to construct a quantum theory in curved spacetime based on such
a quantum equivalence principle, plus some mechanism for propagating thermal
radiation from ``near'' regions out into flatter regions.  (This propagation
process is obviously needed to let Unruh radiation generated at an event
horizon be detected far away as Hawking radiation.)  Yet the propagation of the
thermal radiation does not obviously have to interfere with its generation via
the naive quantum equivalence principle.  Unless this principle can be shown to
be peculiar to the neighbourhood of a horizon, it should apply equally well in
the case of static detectors  outside any spherically symmetric matter
distribution, and at any distance.  These applications of the principle would
not yield the standard results.

Furthermore, the one-to-one association of local directions with orthogonal
field modes, valid in inertial frames in flat space, has been seen to fail for
a constantly accelerating frame.  This does not suggest that the naive quantum
equivalence principle is wrong, but it shows that it will not be as easy to use
as one might hope.  While the Unruh temperature is a genuine property of
acceleration {\it per se}, there is more to acceleration than temperature. The
idea of formulating quantum theory in curved spacetime in terms of observations
made using local detectors would seem to deserve further study.  The technique
of influence functionals should be of use in this regard.

\head{Acknowledgements}

The author gratefully acknowledges valuable discussions with R.C. Myers and A.
Anderson.  This research was supported in part by NSERC of Canada.

\head{Appendix: Rindler cut-off on inertial frequency}

In $D>1$ spatial dimensions, ultra-violet divergent frequency integrals can
appear in accelerating detector problems.  Discarding the high frequency limit
by appeal to the properties of physical detectors is perhaps sophistical, since
any linear acceleration that  produces a measurable Unruh temperature will
destroy any ordinary apparatus[\cite{bell}].  Nevertheless it seems reasonable
that the most  meaningful results are to be extracted from the divergent
integrals by imposing a cut-off at some high frequency $\Gamma$, in the frame
of the accelerating detector.

Given a function $f(x)$ with a Fourier transform $F(k)$, a cut-off on $k$ may
be imposed by changing
$$\eqalign{
f(x)\to f\-\Gamma(x)=&\int_{-\Gamma}^\Gamma\!dk\,F(k) e^{ikx}\cr
                   =&\int_{-\infty}^\infty\!dz\,{\sin\Gamma\!z
                      \over\pi z} f(x+z)\;.}
$$
Thus one may impose a Rindler energy cut-off on the coupling of the detector to
the field, by substituting in \(coup)
$$
e^{ik\-1x\-0(\tau)-i\omega t\-0(\tau)}\to\int_{-\infty}^\infty\!dz{\sin\Gamma\!
z\over\pi z} \,e^{ik\-1x\-0(\tau+z)-i\omega t\-0(\tau +z)}\;.\eqno(recoup)
$$

To obtain the regulated version of $A$ in Equation \(fth1) of Section 5, note
that the delta function $\delta(\tau-\tau')$ forces $\tau=\tau'=\bar\tau$, so
that in the $\bar\tau$ frame
$$\eqalign{
x\-0(\tau+z) &\to {c^2\over a}\cosh{az\over c}\cr
t\-0(\tau+z) &\to {c\over a}\sinh{az\over x}\;.}
$$
Because the $z$ integral is now included in the coupling, \(fth1) is changed
so that
$$
\int_0^\infty {x^2\,dx\over\sqrt{x^2+m^2}}
\to \int_0^\infty {x^2\,dx\over\sqrt{x^2+m^2}}
|C\-\Gamma(x)|^2\;,\eqno(rega)
$$
where
$$\eqalign{
C\-\Gamma(x)&\equiv \int_{-\infty}^\infty\!dz{\sin\Gamma\!z\over\pi z}
e^{i{c^2\over a}(x\cos\theta \cosh{az\over c} -
\sqrt{x^2+m^2}\sinh{az\over c})}\cr
&= \int_{-\infty}^\infty\!dz{\sin\Gamma\!z\over\pi z}
e^{i{c^2\over a}\sqrt{x^2\sin^2\theta + m^2}\sinh\bigl(B(x)
- {az\over c}\bigr)}\;.}\eqno(cgam1)
$$
Here $B(x)$ is defined for convenience, such that
$$
\sinh B(x)\equiv {x\cos\theta
\over\sqrt{x^2\sin^2\theta+m^2}}\;.
$$

Using the identity[\cite{K4}]
$$
e^{i\lambda\sinh\xi} =
{2\over\pi}\int_0^\infty\!d\mu\,K_{i\mu}(\lambda)
\Bigl(\cosh\!{\pi\mu\over2}\cos\mu\xi
+i \sinh\!{\pi\mu\over2}\sin\mu\xi\Bigr)\;,
\eqno(K4)
$$
\(cgam1) may be written
$$\eqalign{
C\-\Gamma(x) &= {2\over\pi}
\int_{-\infty}^\infty\!dz{\sin(\Gamma z)\over\pi z}\cr
&\;\times
\int_0^\infty\!d\mu\,K_{i\mu}({\scc{c^2\over a}}\sqrt{x^2\sin^2\theta +m^2})
\cos\mu\Bigl(B(x)-{az\over c} - {i\pi\over2}\Bigr)\cr
&={2\over\pi}\int_0^{{c\over a}\Gamma}\!d\mu\,
K_{i\mu}({\scc{c^2\over a}}\sqrt{x^2\sin^2\theta +m^2})
\cos\mu\Bigl(B(x) - {i\pi\over2}\Bigr)\;.}\eqno(cgam2)
$$

For fixed order and large argument, the modified Bessel functions of the second
kind have the asymptotic behaviour[\cite{EH2}]
$$
K_{i\mu}(\lambda) \sim \sqrt{\pi\over2\lambda} e^{-\lambda}
\times[1 + {\s O}(\lambda^{-1})]\;.
$$
Therefore the factor $|C\-\Gamma(x)|^2$ does suppress the ultra-violet
divergence in $A$.

\singlespace \references

\refis{K1} I.S. Gradshteyn and I.M. Ryzhik, {\it Table of Integrals, Series,
and Products}, 4th Ed., tr. Alan Geoffrey  (Academic Press;
San Diego, California, 1980), p. 774, eqs. 6.796 -- 2,3

\refis{K2} I.S. Gradshteyn and I.M. Ryzhik, {\it op. cit.},  p. 727, eq. 6.664
-- 6

\refis{fulling} Stephen A. Fulling, \pr D7, 2850, 1973.

\refis{vernon} R.P. Feynman and F.L. Vernon, Jr., \journal Ann. Phys., 24, 118,
1963.

\refis{rindler} W. Rindler, {\it Essential Relativity} (Van Nostrand Reinhold;
New York, 1969), pp. 61-64, 184-195.

\refis{unruh} W.G. Unruh, \pr D14, 870, 1976.

\refis{davies} P.C.W. Davies, \journal J. Phys. A., 8, 609, 1975.

\refis{wald} W.G. Unruh and Robert M. Wald, \pr D29, 1047, 1984.

\refis{coh} A.O. Caldeira and A.J. Leggett, \pr A31, 1067, 1985; W.H.~Zurek,
\pr D26, 1862, 1982; W.G.~Unruh and W.H.~Zurek, \pr D40, 1071, 1989.

\refis{dechis} R. Omn\`es, \journal Rev. Mod. Phys., 64, 339, 1992;
J.J.~Halliwell, \pr D46, 1610, 1992.

\refis{bell} J.S. Bell and J.M. Leinas, \journal Nucl. Phys., B212, 131, 1983.

\refis{jacobson} Theodore Jacobson, \pr D44, 1731, 1991.

\refis{moreau} William Moreau, \journal Am. J. Phys., 60, 561, 1991.

\refis{israel} W. Israel and J.M. Nester, \pl 98A, 329, 1983.

\refis{takagi1} Shin Takagi, \journal Prog. Theor. Phys., 74, 142, 1985.

\refis{takagi2} Shin Takagi, \pl 148B, 116, 1984.

\refis{HPD} K. Hinton, P.C.W. Davies, and J. Pfautsch, \pl 120B, 88, 1983.

\refis{K3} I.S. Gradshteyn and I.M. Ryzhik, {\it op. cit.}, p. 693, eq. 6.576
-- 4; p. 937, eqs. 8.332 -- 1,2; p. 938, eq. 8.335 -- 1.

\refis{dewitt} B.S. DeWitt, in S.W.~Hawking and W.~Israel eds.,
{\it General Relativity: An Einstein Centennial Survey}
(Cambridge University Press; Cambridge, 1979), p. 680.

\refis{K4} I.S. Gradshteyn and I.M. Ryzhik, {\it op. cit.}, p. 774, eq. 6.796
-- 2.

\refis{EH2} A. Erd\'elyi {\it et al.}, {\it Higher Transcendental Functions},
vol. II (McGraw-Hill; New York, 1953), p. 86, sect. 7.13.1 -- (7).

\refis{unruh2} W.G. Unruh, \pr D33, 3573, 1986.

\refis{HPZ} B.L. Hu, Juan Pablo Paz, and Yuhong Zhang, \pr D45, 2843, 1992.

\refis{SRtext} W. Rindler, {\it op. cit.}, p. 108.

\endreferences

\bye